\documentclass[fleqn]{article}
\usepackage{rien-his-styling}
\usepackage{amsmath}
\usepackage{breqn}
\usepackage{svg}
\usepackage{chngcntr}
\usepackage{subcaption}
\usepackage{booktabs}
\usepackage[table]{xcolor} 
\usepackage[backend=biber,style=apa]{biblatex}
\usepackage{lineno}
\usepackage{hyperref}
\usepackage{colortbl} 
\usepackage{float}
\usepackage[acronym, symbols]{glossaries} 
\makeglossaries
\newacronym{am}{AM}{amplitude-modulated}
\newacronym{snr}{SNR}{signal to noise ratio}
\newacronym{assr}{ASSR}{auditory steady-state responses}
\newacronym{eeg}{EEG}{electroencephalogram}
\newacronym{aai}{AAI}{amplitude attenuation index}
\newacronym{rms}{RMS}{root mean square}
\newacronym{dc}{DC}{direct current}
\newacronym{lme}{LME}{linear mixed-effect}
\newacronym{ci}{CI}{confidence interval}
\newacronym{cri}{CrI}{credible interval}
\newacronym{doc}{DoC}{disorders of consciousness}
\newacronym{crs-r}{CRS-R}{Coma Recovery Scale-Revised}
\newacronym{uws}{UWS}{unresponsive wakefulness syndrome}
\newacronym{mcs}{MCS}{minimally conscious state}
\newacronym{tbi}{TBI}{traumatic brain injury}
\newacronym{sca}{SCA}{sudden cardiac arrest}
\newacronym{m}{M}{male}
\newacronym{f}{F}{female}
\newacronym{hl}{HL}{hearing level}
\newacronym{ct}{CT}{computed tomography}
\newacronym{mri}{MRI}{magnetic resonance imaging}
\newacronym{spe}{SpE}{spectral entropy}
\newacronym{ica}{ICA}{independent component analysis}
\newacronym{mse}{MSE}{mean square error}
\newacronym{brm}{BRM}{bayesian regression models}
\newacronym{nt}{NT}{neural tracking}
\newacronym{env}{Env}{envelope}
\newacronym{pe}{PE}{posterior estimate}
\newacronym{srt}{SRT}{speech reception threshold}
\newacronym{ae}{AE}{acoustic onset}
\newacronym{cv}{CV}{cross-validation}
\newacronym{erf}{ERF}{error function}
\newacronym{svr}{SVR}{support vector regression}
\newacronym{shap}{SHAP}{shapley additive explanations}
\newacronym{mae}{MAE}{mean absolute error}
\newacronym{si}{SI}{subject-independent}
\newacronym{ss}{SS}{subject-specific}
\newacronym{L-BFGS}{L-BFGS}{limited-memory Broyden–Fletcher–Goldfarb–Shanno}
\newacronym{pta}{PTA}{Pure-tone audiometry}
\newacronym{eegtech}{EEG}{electroencephalography}
\newacronym{meg}{MEG}{magneticencephalography}
\newacronym{ecog}{ECoG}{electrocorticography}
\newacronym{seeg}{SEEG}{stereo-electroencephalography}
\newacronym{fmri}{fMRI}{functional magnetic resonance imaging}
\newacronym{fnirs}{fNIRS}{functional near-infrared spectroscopy}
\newacronym{pet}{PET}{positron emission tomography}
\newacronym{abr}{ABR}{auditory brainstem response}
\newacronym{caep}{CAEP}{cortical auditory evoked potential}
\newacronym{spl}{SPL}{sound pressure level}
\newacronym{fm}{FM}{frequency modulated}
\newacronym{nrmse}{NRMSE}{normalized root-mean square error}
\newglossaryentry{symb:f}{
  name={\ensuremath{f}},          
  text={\ensuremath{f}},          
  first={frequency (\ensuremath{f})}, 
  description={frequency},
  type=symbols
}

\newglossaryentry{symb:P}{
  name={\ensuremath{P_N(f_i)}},
  text={\ensuremath{P_N(f_i)}},
  first={\ensuremath{P_N(f_i)}},
  description={Normalized power at frequency \ensuremath{f_i}},
  type=symbols
}

\newglossaryentry{symb:au}{
  name={\ensuremath{a.u.}},          
  text={\ensuremath{a.u.}},          
  first={arbitrary units (\ensuremath{a.u.})}, 
  description={arbitrary units},
  type=symbols
}

\newglossaryentry{symb:alpha}{
  name={\ensuremath{\alpha}},          
  text={\ensuremath{\alpha}},          
  description={fixed intercept in a model},
  type=symbols
}
\newglossaryentry{symb:uV}{
  name={\ensuremath{\mu V}},          
  text={\ensuremath{\mu V}},          
  first={micro volt (\ensuremath{\mu V})}, 
  description={micro volt},
  type=symbols
}

\newglossaryentry{symb:infty}{
  name={\ensuremath{\infty}},
  text={\ensuremath{\infty}},
  first={\ensuremath{\infty}},
  description={Infinity; represents an unbounded quantity or limit in mathematics},
  type=symbols
}

\newglossaryentry{symb:T}{
  name={\ensuremath{T}},          
  text={\ensuremath{T}},          
  first={\ensuremath{T}}, 
  description={test statistic},
  type=symbols
}

\newglossaryentry{symb:N}{
  name={\ensuremath{N}},          
  text={\ensuremath{N}},          
  first={\ensuremath{N}}, 
  description={Number of elements (e.g., subjects, bins, components, samples, as defined in context},
  type=symbols
}

\newglossaryentry{symb:p}{
  name={\ensuremath{p}},          
  text={\ensuremath{p}},          
  first={\ensuremath{p}}, 
  description={the probability, under the null hypothesis, of obtaining a result equal to or more extreme than what was observed},
  type=symbols
}

\newglossaryentry{symb:r}{
  name={\ensuremath{r}},
  text={\ensuremath{r}},
  first={\ensuremath{r}},
  description={Pearson correlation coefficient; measures the strength and direction of the linear relationship between two variables},
  type=symbols
}
\newglossaryentry{symb:rs}{
  name={\ensuremath{r_s}},
  text={\ensuremath{r_s}},
  first={\ensuremath{r_s}},
  description={Spearman rank correlation coefficient; measures the strength and direction of the monotonic relationship between two ranked variables},
  type=symbols
}

\newglossaryentry{symb:R2}{
  name={\ensuremath{R^2}},          
  text={\ensuremath{R^2}},          
  first={\ensuremath{R^2}}, 
  description={coefficient of determination},
  type=symbols
}

\newglossaryentry{symb:epsilon}{
  name={\ensuremath{\epsilon}},          
  text={\ensuremath{\epsilon}},          
  first={\ensuremath{\epsilon}}, 
  description={residual error term in a model},
  type=symbols
}

\newglossaryentry{symb:beta}{
  name={\ensuremath{\beta}},          
  text={\ensuremath{\beta}},          
  first={\ensuremath{\beta}}, 
  description={regression coefficients in a model},
  type=symbols
}

\newglossaryentry{symb:t}{
  name={\ensuremath{t}},          
  text={\ensuremath{t}},          
  first={\ensuremath{t}}, 
  description={time variable},
  type=symbols
}

\newglossaryentry{symb:tdist}{
  name={\ensuremath{\mathsf{t}_{\nu}}},  
  text={\ensuremath{\mathsf{t}}},   
  first={\ensuremath{\mathsf{t}}},
  description={Student's t-distribution with \ensuremath{\nu} degrees of freedom},
  type=symbols
}

\newglossaryentry{symb:s}{
  name={\ensuremath{s}},          
  text={\ensuremath{s}},          
  first={seconds (\ensuremath{s})}, 
  description={unit of time, seconds},
  type=symbols
}

\newglossaryentry{symb:ms}{
  name={\ensuremath{ms}},          
  text={\ensuremath{ms}},          
  first={miliseconds (\ensuremath{ms})}, 
  description={unit of time, miliseconds},
  type=symbols
}

\newglossaryentry{symb:Hz}{
  name={\ensuremath{Hz}},          
  text={\ensuremath{Hz}},          
  first={\ensuremath{Hz}}, 
  description={Hertz, unit of frequency},
  type=symbols
}

\newglossaryentry{symb:dB}{
  name={dB},          
  text={dB},          
  first={dB}, 
  description={decibel, logarithmic unit for signal level},
  type=symbols
}

\newglossaryentry{symb:dBA}{
  name={\ensuremath{dBA}},          
  text={\ensuremath{dBA}},          
  first={\ensuremath{dBA}}, 
  description={A-weighted decibels},
  type=symbols
}

\newglossaryentry{symb:dBZ}{
  name={\ensuremath{dBZ}},
  text={\ensuremath{dBZ}},
  first={\ensuremath{dBZ}},
  description={Z-weighted decibels, representing flat frequency weighting across the audible spectrum},
  type=symbols
}

\newglossaryentry{symb:pi}{
  name={\ensuremath{\pi}},          
  text={\ensuremath{\pi}},          
  first={\ensuremath{\pi}}, 
  description={mathematical constant pi},
  type=symbols
}

\newglossaryentry{symb:kHz}{
  name={\ensuremath{kHz}},
  text={\ensuremath{kHz}},
  first={\ensuremath{kHz}},
  description={kilohertz; a unit of frequency equal to $10^3$ Hz},
  type=symbols
}

\newglossaryentry{symb:m}{
  name={\ensuremath{m}},          
  text={\ensuremath{m}},          
  first={sinusoidal modulator (\ensuremath{m})}, 
  description={sinusoidal modulator},
  type=symbols
}

\newglossaryentry{symb:multi}{
  name={\ensuremath{m_M}},          
  text={\ensuremath{m_M}},          
  first={multiplexed modulator (\ensuremath{m_M})}, 
  description={multiplexed modulator},
  type=symbols
}

\newglossaryentry{symb:t2}{
  name={\ensuremath{T^2}},          
  text={\ensuremath{T^2}},          
  first={Hotelling’s statistic (\ensuremath{T^2})}, 
  description={Hotelling’s statistic},
  type=symbols
}

\newglossaryentry{symb:xbar}{
  name={\ensuremath{\bar{x}}},
  text={\ensuremath{\bar{x}}},
  first={\ensuremath{\bar{x}}},
  description={Sample mean; the arithmetic mean calculated from a sample},
  type=symbols
}

\newglossaryentry{symb:g}{
  name={\ensuremath{g}},          
  text={\ensuremath{g}},          
  first={\ensuremath{g}}, 
  description={subject specific decoder model},
  type=symbols
}

\newglossaryentry{symb:G}{
  name={\ensuremath{\mathbf{g}}},          
  text={\ensuremath{\mathbf{g}}},          
  first={\ensuremath{\mathbf{g}}}, 
  description={generic decoder model},
  type=symbols
}

\newglossaryentry{symb:l}{
  name={\ensuremath{l}},
  text={\ensuremath{l}},
  first={\ensuremath{l}},
  description={SNR level of the stimulus},
  type=symbols
}

\newglossaryentry{symb:v}{
  name={\ensuremath{v}},
  text={\ensuremath{v}},
  first={\ensuremath{v}},
  description={Upper limit of integration and input to the error function},
  type=symbols
}

\newglossaryentry{symb:e}{
  name={\ensuremath{e}},
  text={\ensuremath{e}},
  first={\ensuremath{e}},
  description={Euler's number, the base of the natural logarithm, approximately equal to 2.71828},
  type=symbols
}

\newglossaryentry{symb:xi}{
  name={\ensuremath{\xi}},
  text={\ensuremath{\xi}},
  first={\ensuremath{\xi}},
  description={Dummy variable of integration in the error function},
  type=symbols
}

\newglossaryentry{symb:S}{
  name={\ensuremath{S_t}},          
  text={\ensuremath{S_t}},          
  first={actual stimulus envelope (\ensuremath{S_t})}, 
  description={actual stimulus envelope over time \ensuremath{t}},
  type=symbols
}

\newglossaryentry{symb:hatS}{
  name={\ensuremath{\hat{S}_t}},          
  text={\ensuremath{\hat{S}_t}},          
  first={reconstructed stimulus envelope (\ensuremath{\hat{S}_t})}, 
  description={reconstructed stimulus envelope over time \ensuremath{t}},
  type=symbols
}

\newglossaryentry{symb:SD}{
  name={\ensuremath{\mathrm{SD}}},
  text={\ensuremath{\mathrm{SD}}},
  first={\ensuremath{\mathrm{SD}}},
  description={Sample standard deviation; a measure of the dispersion or spread of sample data},
  type=symbols
}

\newglossaryentry{symb:z}{
  name={\ensuremath{z}},
  text={\ensuremath{z}},
  first={\ensuremath{z}},
  description={Standardized statistic for the Wilcoxon signed-rank test},
  type=symbols
}

\newglossaryentry{symb:mu}{
  name={\ensuremath{\mu}},          
  text={\ensuremath{\mu}},          
  first={\ensuremath{\mu}}, 
  description={Mean of a population or distribution},
  type=symbols
}

\newglossaryentry{symb:muhat}{
  name={\ensuremath{\hat{\mu}}},          
  text={\ensuremath{\hat{\mu}}},          
  first={\ensuremath{\hat{\mu}}}, 
  description={Estimated mean (sample mean, or predicted mean from a model)},
  type=symbols
}

\newglossaryentry{symb:phi}{
  name={\ensuremath{\phi}},          
  text={\ensuremath{\phi}},          
  first={\ensuremath{\phi}}, 
  description={precision parameter of the Beta distribution},
  type=symbols
}

\newglossaryentry{symb:u}{
  name={\ensuremath{u}},          
  text={\ensuremath{u}},          
  first={\ensuremath{u}}, 
  description={random intercept in a model},
  type=symbols
}

\newglossaryentry{symb:norm}{
  name={\ensuremath{\mathcal{N}}},          
  text={\ensuremath{\mathcal{N}}},          
  first={\ensuremath{\mathcal{N}}}, 
  description={normal distribution},
  type=symbols
}

\newglossaryentry{symb:sigmasq}{
  name={\ensuremath{\sigma^{2}}},
  text={\ensuremath{\sigma^{2}}},
  first={\ensuremath{\sigma^{2}}},
  description={Variance of a population or distribution},
  type=symbols
}

\newglossaryentry{symb:sigma}{
  name={\ensuremath{\sigma}},
  text={\ensuremath{\sigma}},
  first={\ensuremath{\sigma}},
  description={Standard deviation of a population or distribution},
  type=symbols
}

\newglossaryentry{symb:tau}{
  name={\ensuremath{\tau}},
  text={\ensuremath{\tau}},
  first={\ensuremath{\tau}},
  description={Scale parameter (e.g., spread in t-distribution)},
  type=symbols
}

\newglossaryentry{symb:Prob}{
  name={\ensuremath{\mathbb{P}}},
  text={\ensuremath{\mathbb{P}}},
  first={\ensuremath{\mathbb{P}}},
  description={Probability; $\mathbb{P}(A)$ is the probability that event $A$ occurs.},
  type=symbols
}

\newglossaryentry{symb:D}{
  name={\ensuremath{D}},
  text={\ensuremath{D}},
  first={\ensuremath{D}},
  description={Diagnosis or outcome variable; e.g., $D_j$ denotes the diagnosis for subject $j$.},
  type=symbols
}

\newglossaryentry{symb:C}{
  name={\ensuremath{C}},
  text={\ensuremath{C}},
  first={\ensuremath{C}},
  description={Category; a possible outcome or class in which a diagnosis or variable can fall.},
  type=symbols
}

\newglossaryentry{symb:kappa}{
  name={\ensuremath{\kappa}},
  text={\ensuremath{\kappa}},
  first={\ensuremath{\kappa}},
  description={Threshold parameter (intercept) for each cumulative probability boundary in ordinal regression models},
  type=symbols
}

\newglossaryentry{symb:i}{
  name={\ensuremath{i}},
  text={\ensuremath{i}},
  first={\ensuremath{i}},
  description={Index for observations or data points},
  type=symbols
}

\newglossaryentry{symb:j}{
  name={\ensuremath{j}},
  text={\ensuremath{j}},
  first={\ensuremath{j}},
  description={Index for subjects, individuals, or groups},
  type=symbols
}

\newglossaryentry{symb:nt}{
  name={\ensuremath{NT}},
  text={\ensuremath{NT}},
  first={\ensuremath{NT}},
  description={Fixed effect for neural tracking in the statistical model},
  type=symbols
}

\newglossaryentry{symb:spe}{
  name={\ensuremath{SpE}},
  text={\ensuremath{SpE}},
  first={\ensuremath{SpE}},
  description={Fixed effect for spectral entropy in the statistical model},
  type=symbols
}

\newglossaryentry{symb:x}{
  name={\ensuremath{x}},
  text={\ensuremath{x}},
  first={\ensuremath{x}},
  description={Observation of a variable (e.g., neural tracking, spectral entropy)},
  type=symbols
}

\newglossaryentry{symb:w}{
  name={\ensuremath{w}},
  text={\ensuremath{w}},
  first={\ensuremath{w}},
  description={The slope parameter for the error function},
  type=symbols
}

\title{A Multi-decoder Neural Tracking Method for Accurately Predicting Speech Intelligibility}

\name{Rien Sonck$^{1}$, Bernd Accou$^{1}$, Tom Francart$^{1}$, Jonas Vanthornhout$^{1}$}
\address{
  $^1$ ExpORL, Dept. of Neurosciences, KU Leuven, Leuven, Belgium
}   
\email{jonas.vanthornhout@kuleuven.be}

\begin{document}
\maketitle
\begin{abstract}

\textit{Objective}: \gls{eeg}-based methods have shown promise to predict behaviorally measured speech intelligibility, but accuracy and robustness generally fall short of standard behavioral test–retest procedures, which often yield test-retest differences of less than 1~\gls{symb:dB}. We introduce the multi-decoder method to predict \glspl{srt} from \gls{eeg} recordings of speech listening. This method is designed for populations unable to perform behavioral tests, allowing an objective assessment of speech intelligibility in various listening environments and informing clinical decision-making, hearing aid fitting or assessment of patients with \gls{doc}. 

\textit{Approach}: The multi-decoder method aggregates information from hundreds of decoders, each trained using different combinations of speech features and EEG preprocessing configurations to compute neural tracking \gls{nt}, quantifying each decoder’s ability to reconstruct speech signals from \gls{eeg}. To train and test the decoders, we used data from 39 participants (7 men, 32 women), aged 18 to 24 years, with a total \gls{eeg} recording duration of 29 minutes per participant. Each participant first completed a behavioral task to estimate their \gls{srt}. Subsequently, \gls{eeg} recordings were acquired while they listened to speech presented at six different \glspl{snr}, as well as a narrated story presented in quiet. The \gls{nt} values from all decoders were then combined into a single high-dimensional feature vector per subject. Then a \gls{svr} model was trained to learn the relationship between these \gls{nt} vectors and the behavioral \glspl{srt}, with the goal of predicting individual \glspl{srt} from the \gls{nt} vectors.

\textit{Main result}: Predictions using the multi-decoder method correlated significantly (r=0.647, $p<0.001$) with behavioral \glspl{srt}(NRMSE =0.19); all differences stayed below 1~\gls{symb:dB}. \gls{shap} analysis indicated that while all features contributed, the theta and delta frequency bands, early lags, and subject-specific decoders had a slightly greater influence on model predictions. Furthermore, we demonstrated that leveraging pretrained \gls{si} decoders allows reducing \gls{eeg} data collection for new participants to just 15 minutes, comprising 3 minutes of story listening and 12 minutes across six \gls{snr} conditions with 40 matrix sentences each—without sacrificing prediction accuracy. 

\textit{Significance}: The multi-decoder method introduced in this study outperforms previous \gls{eeg}-based approaches. Future research should validate the method in more diverse clinical populations, including those with hearing loss.

\end{abstract}
\glsresetall

\section{Introduction}

Assessing speech intelligibility, the extent to which a listener can understand spoken words and sentences, is a fundamental aspect of evaluating human auditory function. A widely used metric for this purpose is the \glsreset{srt}\gls{srt}, defined as the \glsreset{snr}\gls{snr} at which a listener can correctly repeat 50\% of the presented speech material. \glspl{srt} are useful for evaluating hearing-impaired listeners, with or without hearing aids, in challenging listening environments \parencite[]{verschuureEffectHearingAids1992, festenSpeechreceptionThresholdNoise1986, plompSignaltonoiseRatioModel1986}, as well as for assessing the effectiveness of specific hearing aid algorithms \parencite[]{zaarSpectrotemporalModulationTest2024, vandijkhuizenEffectVaryingAmplitudefrequency1989}.

Behavioral \gls{srt} tasks, have been developed for various languages to assess speech intelligibility in noise. One early example is the Swedish matrix test \parencite[]{hagermanSentencesTestingSpeech1982}, which has since inspired adaptations in multiple languages, including Dutch and Flemish \parencite[]{lutsDevelopmentNormativeData2014,houbenDevelopmentDutchMatrix2014}. However, these behavioral assessments require active participation, which can be challenging or unfeasible when testing children or certain clinical groups. As a result, neuroimaging-based methods have gained interest in measuring speech intelligibility. 

Neuroimaging studies have employed backward (i.e. decoder) and forward (i.e. encoder) models of neural tracking \parencite[for a review][]{gillisNeuralTrackingDiagnostic2022}. These models seek to establish a linear mapping between the features of the speech stimulus (e.g., the envelope, the acoustic onsets, and the phonemes) and the corresponding brain responses. Backward models reconstruct the speech feature from the multichannel \gls{eeg} signal, whereas forward models predict the \gls{eeg} signal for each channel from the stimulus feature. For backward models, the Pearson correlation between the reconstructed feature and the actual feature is called the \glsreset{nt}\gls{nt} value. For forward models, this is the Pearson correlation between the reconstructed \gls{eeg} signal and the actual \gls{eeg} signal. 

Perception of the speech envelope is important for speech intelligibility \parencite[]{peelleNeuralOscillationsCarry2012, shannonSpeechRecognitionPrimarily1995, smithChimaericSoundsReveal2002}, which is further supported by studies showing a strong correlation between behaviorally measured speech intelligibility and neural tracking of the speech envelope \parencite[]{vanthornhoutSpeechIntelligibilityPredicted2018}. 
However, while broadband envelope perception is necessary for intelligibility, it is not always sufficient; spectro-temporal fine structure also plays an important role, especially in noisy or challenging listening conditions \parencite[]{dingRobustCorticalEntrainment2014, lorenziSpeechPerceptionProblems2006, mooreRoleTemporalFine2008}. Decoding of the speech envelope is further influenced by factors such as spectral degradation \parencite[]{kosemNeuralTrackingSpeech2023, karunathilakeNeuralTrackingMeasures2023b}, speech rate \parencite[]{verschuerenSpeechUnderstandingOppositely2022}, attention \parencite[]{vanthornhoutEffectTaskAttention2019} and binaural unmasking \parencite[]{dieudonneNeuralTrackingSpeech2025}. Notably, neural tracking of the envelope also occurs for foreign or unintelligible speech \parencite[]{zouAuditoryLanguageContributions2019, songCrosslinguisticPerceptionContinuous2016, reetzkeNeuralTrackingSpeech2021a, ortizbarajasOriginsDevelopmentSpeech2021, gillisHeardUnderstoodNeural2023}; nevertheless, the envelope remains an important acoustic feature for speech understanding.

While the speech envelope is an important feature for speech intelligibility, acoustic onsets also play a role in supporting speech understanding. Acoustic onsets are derived from the half-wave rectified first derivative of the speech envelope. Acoustic onsets are important for segmenting and organizing speech and carry essential phonetic information \parencite[]{kluenderSensitivityChangePerception2003, brodbeckRapidTransformationAuditory2018}. Enhancing acoustic onsets in cochlear implants has been shown to improve speech intelligibility \parencite[]{koningPotentialOnsetEnhancement2012, koningSpeechOnsetEnhancement2016}. However, like the envelope, acoustic onsets are processed by the auditory system regardless of speech intelligibility \parencite[]{karunathilakeNeuralTrackingMeasures2023b}.

Currently, only a limited number of studies have proposed objective \gls{eeg}-based methods for estimating speech intelligibility. \textcite{iotzovEEGCanPredict2019} implemented a hybrid approach that combined forward and backward modeling to assess speech intelligibility; however, they did not attempt to estimate \glspl{srt}. They found that speech detection improved with intelligibility, as reflected by increased \gls{eeg}–speech correlation. Approaches specifically designed to predict the \gls{srt} from \gls{eeg} data began with the work of \textcite{vanthornhoutSpeechIntelligibilityPredicted2018}, who employed a backward modeling approach. In this approach, neural tracking values were computed across a range of \gls{snr} conditions, and a sigmoid function was fitted to these values, analogous to fitting a psychometric curve in behavioral experiments. The midpoint of the fitted curve, referred to as the correlation threshold, was used as an estimate of the behavioral \gls{srt}, yielding a correlation of 0.69 between them, with a median absolute difference of 1.7~\gls{symb:dB}, and a standard deviation of 3.17~\gls{symb:dB} However, the approach was not successful for all participants: in 21\% of the cases (5 out of 24 subjects), the correlation threshold could not be reliably determined. An alternative approach using forward modeling was investigated by \textcite{lesenfantsPredictingIndividualSpeech2019a}, who evaluated a variety of speech characteristics, including envelope, spectrogram, phonemes, phonetic characteristics and a combined feature set termed FS (phonetic characteristics + phonemes of the spectrogram). The FS feature provided the best performance and did not require sigmoid fitting. Instead, they introduced the FS-zero crossing, defined as the \gls{snr} at which the neural tracking values transitioned from negative to positive. This measure was effective for predicting behavioral \glspl{srt} and achieved a mean absolute difference of 0.97~\gls{symb:dB}.
Another study using forward modeling was conducted by \textcite{borgesSpeechReceptionThreshold2025}, who used the envelope feature and also applied sigmoid fitting. Their method achieved a median difference of 0.38~\gls{symb:dB} with a standard deviation of 1.45~\gls{symb:dB}, but failed to reliably fit 9\% of participants (2 out of 22). \textcite{munckePredictionSpeechIntelligibility2022} predicted behavioral \glspl{srt} by combining forward models with a root mean square approach, eliminating the need for sigmoid fitting and achieving a mean difference of 1.2~\gls{symb:dB}. 

While most work on deep learning and speech intelligibility does not directly estimate \glspl{srt}, one notable exception is the study by \textcite{accouPredictingSpeechIntelligibility2021}. In this work, a dilated convolutional neural network was trained using a match–mismatch paradigm, and sigmoid fitting was used to estimate \glspl{srt} from classification accuracy across \glspl{srt}. This method achieved a correlation of 0.59 with the behavioral \glspl{srt}, a median absolute difference of 3.64~\gls{symb:dB}, and a standard deviation of 1.68~\gls{symb:dB}.

Despite recent advances, two main challenges remain. First, many existing approaches still struggle to provide reliable \gls{srt} predictions at the individual level. In contrast, the behavioral matrix test consistently delivers accurate \gls{srt} measurements, with studies reporting very low test–retest variability—median differences around 0.4~\gls{symb:dB} \parencite[]{boonMulticenterEvaluatieValidatie2014} and a standard deviation of 0.7~\gls{symb:dB} \parencite[]{decruySelfAssessedBekesyProcedure2018}. While \gls{eeg}-based methods are advancing and gradually closing this accuracy gap, they have yet to reach the robustness of behavioral benchmarks. Second, comparing \gls{srt} prediction performance across studies is difficult due to variations in speech material and differences in the distribution of behavioral \glspl{srt} among participants.

In this work, we propose a new method to tackle the first challenge: the multi-decoder method. The key innovation lies in combining information across a large set of decoders, each trained on differing configurations of stimulus features (envelope and acoustic onsets), integration windows (ranging from 0 to 500~\gls{symb:ms}), frequency bands (delta, theta, and broadband), and both subject-specific and subject-independent decoder configurations. This removes the constraint of having to select these parameters in advance for all subjects. Furthermore, our method replaces traditional sigmoid fitting by combining the error function with a \gls{svr} approach, enabling robust and accurate \gls{srt} predictions across all participants. Additionally, we analyze how much \gls{eeg} data is necessary to maintain optimal the model performance. Specifically, we want to know if pretrained subject-independent decoders can reduce the amount of \gls{eeg} data we require from new participants to keep the same model performance in predicting their \gls{srt}—a crucial step toward practical application in clinical settings.

For the second challenge: to compare \gls{srt} prediction performance across speech material and \gls{srt} distributions, we advocate for more standardized reporting across studies. To this end, we propose using the \gls{nrmse} as a metric, as it accounts for variability in the data distribution and enables more meaningful comparisons of \gls{srt} prediction performance.

\section{Materials and methods}

\subsection{Participants}
A total of 39 participants (7 men, 32 women), aged between 18 and 24 years (\gls{symb:xbar} = 21, \gls{symb:SD} = 1.8), were included from the dataset described by \cite[]{accouSparrKULeeSpeechEvokedAuditory2024}. All participants had hearing thresholds within the normal to slight hearing loss range, as confirmed by pure-tone air conduction audiometry. Most thresholds were $\leq$ 20 \gls{symb:dB} \gls{hl} across 125–8000 \gls{symb:Hz}, with a small number between 21 to 30 \gls{symb:dB} \gls{hl}. Prior to participation, all subjects provided written informed consent of the protocol approved by the Medical Ethics Committee UZ KU Leuven/Research (KU Leuven, Belgium; reference S57102).

\subsection{Experimental Procedure}
\label{sec:experimental_procedure}

\begin{figure}[htbp]
  \centering
  \includegraphics[width=0.45\textwidth]{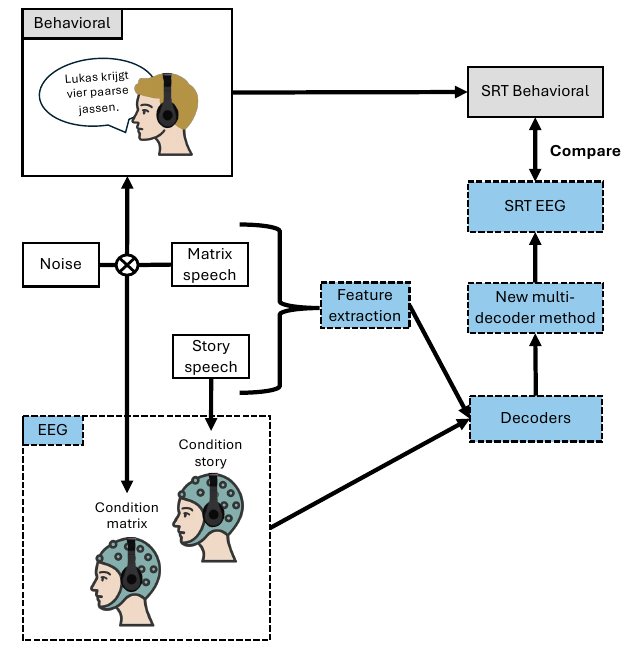}
    \caption[Overview of the experimental procedure]{
    \textbf{Overview of the experimental procedure.}
   Each participant completed three main tasks: a behavioral task in which they listened to matrix sentences masked by speech-weighted noise and repeated them to determine their behavioral \gls{srt}; followed by two \gls{eeg} tasks involving passive listening—one to a narrated story, and the other to matrix sentences presented with varying levels of speech-weighted noise.
    Feature extraction focused on the envelope and acoustic onsets of the clean stimuli. Both the stimulus features and the \gls{eeg} recordings from the narrated story and matrix sentences were then used to train and test hundreds of different decoder configurations. These decoders were subsequently combined using our proposed multi-decoder method, which integrates information from all decoders to predict the \gls{srt}, allowing comparison with the behavioral \gls{srt}.
    }
  \label{fig:fig1}
\end{figure}

Each participant completed a series of tasks as detailed by \cite[]{accouSparrKULeeSpeechEvokedAuditory2024}. For the present study, we analyzed data from one behavioral task (hereafter, the matrix behavioral task) and two \gls{eeg} tasks: the story task and the matrix \gls{eeg} task. All auditory stimuli were presented binaurally at 62~\gls{symb:dBA}. \gls{eeg} data were recorded from 64 scalp electrodes at a sampling rate of 8192~\gls{symb:Hz}. For additional details on the \gls{eeg} acquisition, see \cite[]{accouSparrKULeeSpeechEvokedAuditory2024}. See \textit{Figure \ref{fig:fig1}} for an overview of the experimental procedure. 

\subsubsection{Matrix Behavioral Task}
The Flemish matrix behavioral task \parencite[]{lutsDevelopmentNormativeData2014} used an adaptive procedure to determine each participant’s \gls{srt}. Each test list consisted of 20 standardized sentences (one matrix list), presented against stationary speech-weighted noise fixed at 65~\gls{symb:dB} SPL. For full task details, see \textcite{lutsDevelopmentNormativeData2014}.

\subsubsection{Story Task}
In the story task, participants listened to a 15-minute Flemish-language story, "Milan",  narrated and written by Stijn Vranken, a male speaker, while their \gls{eeg} was recorded.

\subsubsection{Matrix EEG Task}
During the matrix \gls{eeg} task, participants listened attentively to matrix sentences presented in speech-weighted noise at fixed \glspl{snr}: $-12.5$, $-9.5$, $-6.5$, $-3.5$, $-0.5$, and $2.5$~\gls{symb:dB}, as well as a silent condition. These are collectively referred to as the \gls{snr} conditions. Participants were told to listen passively. Each \gls{snr} condition comprised two matrix lists, totaling 40 sentences with brief pauses between them. All participants completed every \gls{snr} condition, and the order of conditions was randomized per participant. Each condition lasted approximately two minutes, resulting in a total task time of around 14 minutes.

\subsection{Speech feature extraction}
\label{sec:feature_extraction}

The speech signals were sampled at 48~\gls{symb:kHz}. The speech envelope and acoustic onset were extracted from clean versions of both the story and the matrix stimuli.

\subsubsection{Envelope}
To extract the temporal \gls{env}, the audio was processed with a gammatone filter bank using 28 subbands, with center frequencies from 50~\gls{symb:Hz} until 5000~\gls{symb:Hz}, followed by power-law compression: each subband envelope was calculated by taking the absolute value of each sample and raising it to the power of 0.6. The final broadband envelope was obtained by averaging across all subbands, using the same filter bank configuration as described by \textcite{biesmansAuditoryInspiredSpeechEnvelope2017, vanthornhoutSpeechIntelligibilityPredicted2018}. 

\subsubsection{Acoustic onsets}
\Glspl{ae} were computed as the first derivative of the speech envelope (that is, the difference between consecutive envelope samples) and applying half-wave rectification, which retains only positive changes. 

\subsection{Signal preprocessing}
\label{sec:signal_preprocessing}
The \gls{eeg} data were first high-pass filtered using a first-order zero-phase Butterworth filter with a cutoff frequency of 0.5~\gls{symb:Hz}. The data were then downsampled to 1024~\gls{symb:Hz}. Eye blink artifacts were removed using a multichannel Wiener filter \parencite[]{somersGenericEEGArtifact2018}, and any bad channels identified during the session \gls{eeg} were linearly interpolated. The signals were re-referenced to the common average.

Both \gls{eeg}, the envelope of the stimuli, and \gls{ae} of the stimuli were downsampled to 64~\gls{symb:Hz}. Three least squares band pass filters (order 2000), each designed with a stopband attenuation of 80~\gls{symb:dB} and a passband ripple of 1~\gls{symb:dB}, were applied separately to extract the delta (0.5–4~\gls{symb:Hz}), theta (4–8~\gls{symb:Hz}), and broadband (0.5–30~\gls{symb:Hz}) frequency bands. Finally, all \gls{eeg} channels and speech features were z-score normalized across the whole recording.

\subsection{Decoders}

To assess how well the brain tracks or aligns with speech features, specifically the envelope and acoustic onsets of both the narrated story and matrix sentences, we used a decoder (i.e., backward model). This approach reconstructs speech features from neural data and compares them with the original speech features, providing a measure of neural tracking quantified by the Pearson correlation between the actual and reconstructed features \parencite[for a comprehensive overview, see][]{gillisNeuralTrackingDiagnostic2022}. Once trained, the decoder can be applied to unseen neural data to reconstruct the corresponding stimulus feature. The resulting correlation coefficient serves as the neural tracking value: Higher correlations indicate a more accurate reconstruction of the stimulus feature from neural data, which can be interpreted as a marker of speech intelligibility for the participant \parencite[]{vanthornhoutSpeechIntelligibilityPredicted2018}.

\subsubsection{Decoder configurations}

A decoder configuration is defined by a set of parameters used to set up the decoder training and testing, including the \gls{eeg} task (story vs. matrix), the speech feature (envelope vs. acoustic onsets), the integration window (27 windows), frequency bands (theta, delta, and broadband) and the type of decoder (subject-independent vs. subject-specific). Resulting in a total of 648 unique decoder configurations. Each decoder configuration is used to train and test seven decoders, one for each \gls{snr} condition: $-12.5$, $-9.5$, $-6.5$, $-3.5$, $-0.5$, $2.5$~\gls{symb:dB}, and silence, resulting in a total of 4536 decoders. For each decoder we used ridge regression to estimate the optimal set of weights \parencite[]{machensLinearityCorticalReceptive2004} that maps the \gls{eeg} data to the speech feature of interest, the speech envelope or the acoustic onsets. Although decoders can be trained on data from either the matrix or story \gls{eeg} tasks, all decoders are exclusively tested on the matrix \gls{eeg} task, as performance on this task provides the necessary information to predict the \gls{srt}.

The \gls{eeg} task is defined in Section~\ref{sec:experimental_procedure}, the speech features in Section~\ref{sec:feature_extraction}, and the frequency bands in Section~\ref{sec:signal_preprocessing}. Below, we provide details on the post-integration window and the type of decoders.

 \subsubsection{Post-stimulus integration window}

Because neural responses to sound occur with varying delays due to auditory processing latencies, the \gls{eeg} data was expanded with multiple time-lagged versions of itself. This post-stimulus integration window was systematically varied from 0 to 500~\gls{symb:ms}. Sets of integration windows were constructed by starting at 0 and incrementally increasing the maximum time lag: first from 0 to 5 samples, then from 0 to 6 samples, and so on. At a sampling rate of 64~\gls{symb:Hz}, these correspond to lag ranges of 0–78~\gls{symb:ms}, 0–94~\gls{symb:ms}, 0–109~\gls{symb:ms}, etc., with each additional window extending the maximum lag by approximately 15 to 16~\gls{symb:ms} until 500~\gls{symb:ms}. Resulting in a total of 27 windows. 

\subsubsection{Decoder training}

\begin{figure*}[htbp]
  \centering
  \includegraphics[width=0.9\textwidth]{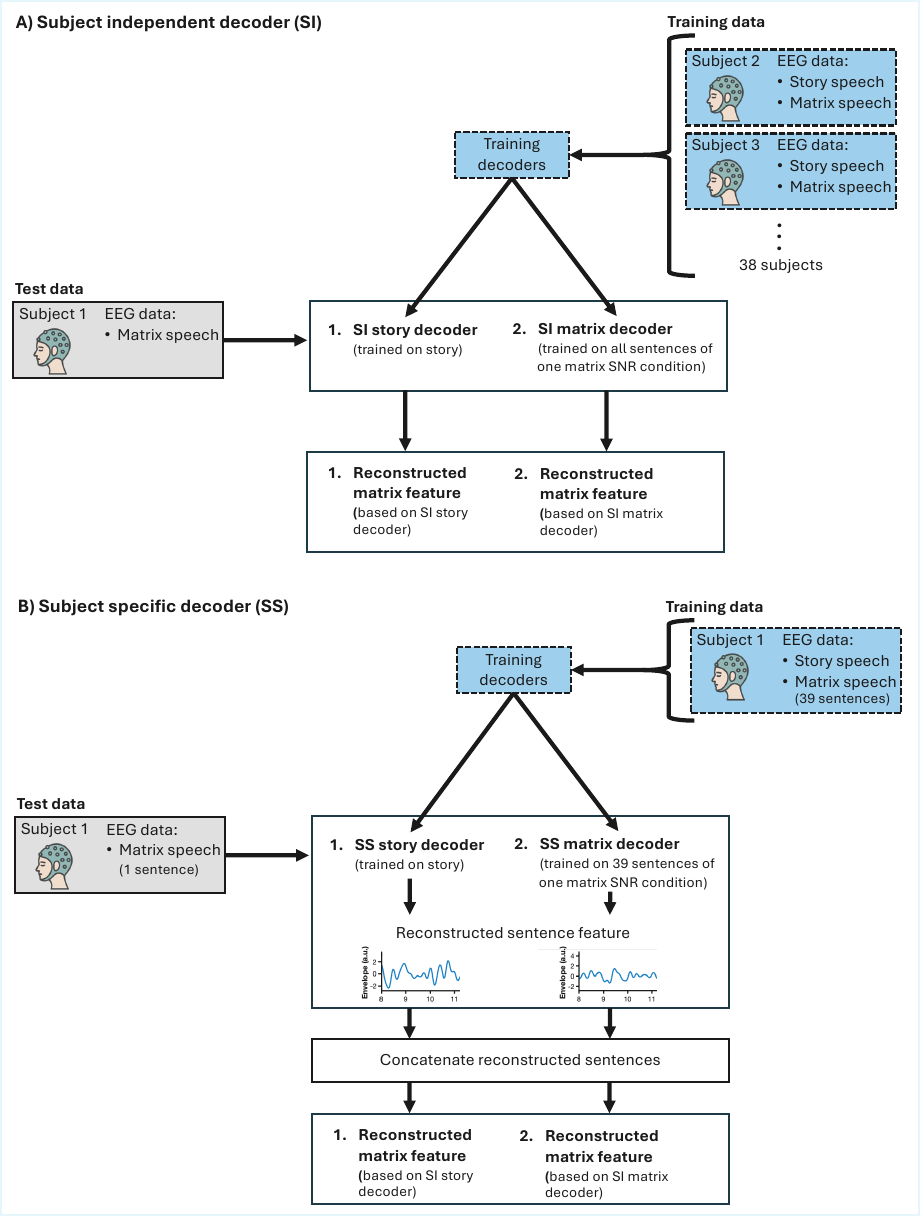}
    \caption[Subject-independent and subject-specific decoders]{
    \textbf{Subject-independent and subject-specific decoders.} (A) For the \gls{si} decoder, leave-one-subject-out cross-validation is applied; this panel depicts the process for a single subject, representing one fold of the cross-validation. Two \gls{si} decoders are shown: one trained on \gls{eeg} data from the story-listening task (\gls{si} story decoder) and another trained on \gls{eeg} data from a specific \gls{snr} condition of the matrix listening task (\gls{si} matrix decoder). Each \gls{snr} condition includes two matrix sentence lists, totaling 40 sentences. (B) For the \gls{ss} decoder, leave-one-out cross-validation was performed on matrix sentences within a single \gls{snr} condition. Similar to the \gls{si} decoders, two \gls{ss} decoders were trained: one on \gls{eeg} story data (\gls{ss} story decoder) and one on \gls{eeg} data from 39 of the 40 matrix sentences (\gls{ss} matrix decoder). After reconstructing the features for all 40 sentences across the cross-validation folds, the reconstructed sentence features were concatenated into a single feature vector.
    }
  \label{fig:fig2}
\end{figure*}

We used \gls{si} and \gls{ss} decoders. The \gls{si} decoder was trained and evaluated across all participants using a leave-one-subject-out \gls{cv} approach (see \textit{Figure \ref{fig:fig2}A}). In this procedure, the data each participant served once as test data, while data from the remaining participants were used for training. The decoder was trained using \gls{eeg} data from either the story task or from a single \gls{snr} condition of the matrix task. As noted previously, each decoder was tested only on one corresponding \gls{snr} condition of the matrix task \gls{eeg} data.

The \gls{ss} decoder was trained and tested using data from a single participant (see \textit{Figure \ref{fig:fig2}B}). Two training configurations were used. In the first, the decoder was trained on the participant’s \gls{eeg} data from the story task and tested on one \gls{snr} condition of that participant’s matrix task. In the second configuration, the decoder was both trained and tested on data from the same matrix task. As each \gls{snr} condition contained 40 matrix sentences, a leave-one-out sentence-wise cross-validation was applied: 39 sentences were used to train the decoder, and the remaining sentence was used for testing and feature reconstruction. This procedure was repeated until each sentence had been reconstructed once. The reconstructed stimulus features from all 40 test sentences were then concatenated—excluding silent intervals—and correlated with the corresponding actual stimulus features, yielding a single \gls{nt} value.

\subsection{Predicting the SRT}

\subsubsection{Adjusted neural tracking values}

\begin{figure*}[htbp]
  \centering
  \includegraphics[width=1.0\textwidth]{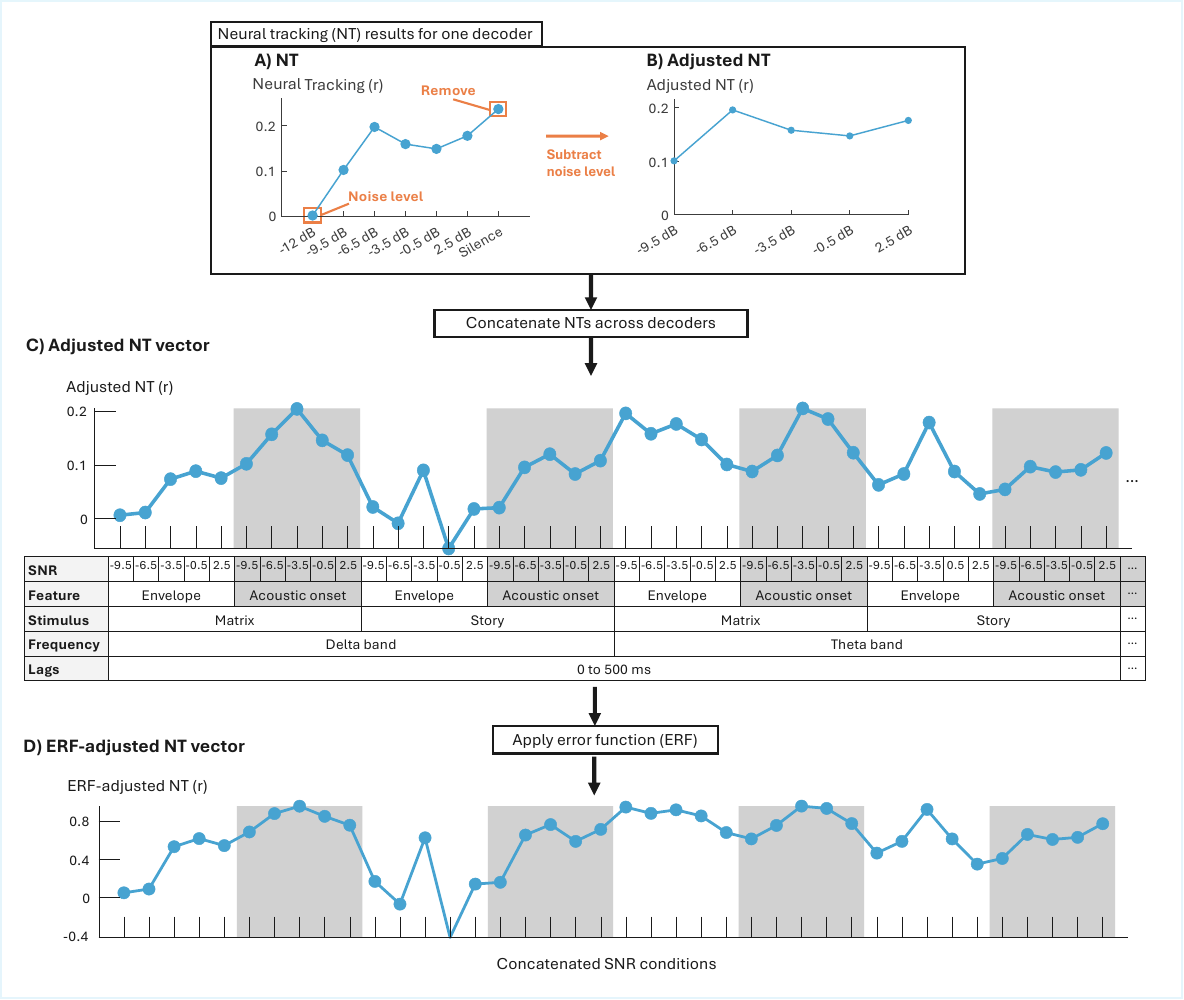}
    \caption[ERF-adjusted neural tracking vectors]{
    \textbf{ERF-adjusted neural tracking vectors.} 
    (A) Displays \gls{nt} values for seven \glspl{snr} conditions using a specific decoder configuration (e.g., theta-band, story listening, stimulus envelope, lags up to 250 ms). Seven decoders—one per \gls{snr} condition—were trained and tested using the same decoder configuration, resulting in seven neural tracking values corresponding to each condition. The \gls{nt} value at –12.5~\gls{symb:dB} \gls{snr} serves as the noise baseline, and the silence condition is excluded from further analysis. (B) Adjusted \gls{nt} values are computed by subtracting the noise baseline (–12.5~\gls{symb:dB} \gls{snr}) from the \gls{nt} values at the remaining \gls{snr} conditions. (C) Each decoder configuration produces its own set of adjusted \gls{nt} values across the \glspl{snr}. These \gls{nt} values from all decoder configurations are concatenated into a single comprehensive adjusted \gls{nt} vector. The alternating grey and white backgrounds highlight the adjusted \gls{nt} values corresponding to the same decoder configuration across different \glspl{snr}. For clarity, the decoder type parameter is not shown in this illustration. (D) Each subject's adjusted \gls{nt} vector is transformed into an \gls{erf}-adjusted vector using the \glsreset{erf}\gls{erf}.}
    \label{fig:fig3}
\end{figure*}

 Each decoder configuration is used to train and test seven decoders, one for each \gls{snr} condition, resulting in seven neural tracking (\gls{nt}) values: $-12.5$, $-9.5$, $-6.5$, $-3.5$, $-0.5$, $2.5$~\gls{symb:dB}, and silence, shown in \textit{(Figure \ref{fig:fig3}A)}. The silence condition was excluded from further analysis, as neural responses to speech in noise differ fundamentally from those to speech in silence \parencite[]{dilibertoIndexingCorticalEntrainment2017}. The $-12.5$~\gls{symb:dB} \gls{snr} condition was chosen as a noise baseline, because it falls well below the \gls{srt} for all participants. \gls{nt} values at this baseline were subtracted from the \gls{nt} values at the other \glspl{snr}, resulting in five adjusted \gls{nt} values for SNRs of $-9.5$, $-6.5$, $-3.5$, $-0.5$, and $2.5$~\gls{symb:dB}, see \textit{Figure \ref{fig:fig3}B}.

\subsubsection{Feature: ERf-adjusted NT vectors}

As each decoder configuration resulted in five adjusted \gls{nt} values across \glspl{snr}, there are a total of 3240 adjusted \gls{nt} values.  For each subject, all these adjusted \gls{nt} values are concatenated into a single, \gls{nt} vector (see \textit{Figure \ref{fig:fig3}C}). The \gls{nt} vector is  then transformed using the \glsreset{erf}\gls{erf} \parencite[]{abramowitzHandbookMathematicalFunctions1972} (see \textit{Figure \ref{fig:fig3}D}), resulting in \gls{erf}-adjusted \gls{nt} vectors. The \gls{erf} replaces the sigmoid fitting, transforming the curve into a smooth S-shape, similar to a psychometric function, but without requiring subject-specific curve fitting. 

\subsubsection{Model: support vector regression}
Each participant was associated with an \gls{erf}-adjusted \gls{nt} vector. To predict the corresponding \gls{srt} values from these vectors,  a linear \gls{svr} model \parencite[]{smolaTutorialSupportVector2004} was used with ridge regularization and an efficient optimization algorithm based on \glsentryshort{L-BFGS} \parencite[]{hoerlRidgeRegressionBiased1970,hoLargescaleLinearSupport2012}. This approach models the linear relationship between the \gls{erf}-adjusted \gls{nt} vectors and the \gls{srt} outcomes, while controlling for overfitting through regularization.

Model performance was evaluated using a nested leave-one-out \gls{cv} procedure to ensure unbiased hyperparameter tuning and testing. In the outer loop, one subject was held out as an test case, while the remaining subjects constituted the training set. Within this training set, an inner leave-one-out \gls{cv} was performed, where one subject was used as a validation set to optimize the hyperparameter \gls{symb:sigma}, which controls the steepness of the S-shaped error function curve. This curve models the expected neural tracking performance \(\Psi(l)\) as a smooth sigmoid function of the stimulus \gls{snr} level \(l\), centered at \(\mu\), the threshold or central  \gls{snr} value:

\begin{equation}
\Psi(l) = \frac{1}{2} \left[ 1 + \mathrm{erf}\left( \frac{l - \mu}{\sigma \sqrt{2}} \right) \right],
\label{eq:mat-eq2}
\end{equation}

This \gls{erf} transformation was applied elementwise to each decoder configuration (grey and white background squares on \textit{(Figure \ref{fig:fig3}C}) of the \gls{nt} vectors, following Equation~\eqref{eq:mat-eq2}. Importantly, the steepness parameter \gls{symb:sigma} was fitted globally across all decoder configurations of the vector. This means that while every decoder configuration was individually transformed by the \gls{erf} function, the same fitted value of \gls{symb:sigma} was used consistently across all decoder configurations.

Within each fold of the inner leave-one-out cross-validation loop, a different candidate  value (ranging from 1 to 10) was tested. For each candidate, the adjusted \gls{nt} vector was transformed into the \gls{erf}-adjusted \gls{nt} vector. Subsequently, a support vector regression (\gls{svr}) model was trained on the inner training set and evaluated on the validation subject. The optimal \gls{symb:sigma} was determined by minimizing the \gls{mae} between the true \gls{srt} ($y_i$) and the predicted \gls{srt} ($\hat{y}_i$) across all subjects in the inner loop:
\begin{equation}
MAE = \frac{1}{N} \sum_{i=1}^{N} |y_i - \hat{y}_i|
\label{eq:mat-eq3}
\end{equation}
where $N$ is the number of subjects in the inner loop.
The \gls{symb:sigma} that produced the lowest \gls{mae} was then selected, and the \gls{erf} with this optimal \gls{symb:sigma} was applied to the adjusted \gls{nt} vectors of all subjects in the training set. The \gls{svr} was then retrained on the entire outer training set and finally tested on the held-out test subject, resulting in a predicted \gls{srt} for that subject.

\subsubsection{Model performance}

Model performance was evaluated by correlating the predicted \gls{srt} values with the corresponding behavioral \gls{srt} measurements. To determine whether the observed correlation was statistically significant, we constructed a null distribution of model performance (\textit{Figure \ref{fig:figure7}}). This null distribution was generated by randomly shuffling the behavioral \glspl{srt} across participants, thereby breaking the relationship between the behavioral \glspl{srt} and the \gls{erf}-adjusted \gls{nt} vectors. Using these shuffled data, we trained the \gls{svr} model anew using leave-one-out cross-validation  to obtain a set of predicted \glspl{srt}, which were then correlated with the behavioral \glspl{srt}. This procedure was repeated 1000 times to yield a null distribution of correlation values. Because correlation coefficients are bounded between -1 and +1 and therefore non-normally distributed, we applied a Fisher Z-transformation to the null distribution. The resulting Z-transformed values approximate a normal distribution, enabling us to fit a Gaussian curve to the null data and perform statistical testing. Specifically, we calculated a two-tailed p-value by comparing the Fisher Z-transformed observed correlation to the null distribution.

Another way to evaluate model performance is by calculating the absolute difference between the predicted \gls{srt} and the behavioral \gls{srt}, as demonstrated by \cite{lesenfantsPredictingIndividualSpeech2019a}. This difference should ideally be less than 1~\gls{symb:dB}, comparable to the test-retest variability observed in the behavioral matrix task itself.

\subsubsection{Evaluating the effect of data reduction on model performance}

In this analysis, we aim to determine the minimum amount of \gls{eeg} data required to maintain model performance comparable to using the full 27 minutes per subject (excluding the unused silence condition). To do this, we systematically reduced data from both the story and matrix \gls{eeg} tasks.

For the matrix \gls{eeg} task, we progressively reduced the number of matrix sentences per \gls{snr} condition from 40 down to 5, decreasing in steps of 5 sentences while keeping the full 15 minutes of story \gls{eeg} data intact. This reduction affects both training and testing datasets involving matrix sentences.

For the story \gls{eeg} task, we decreased the duration from 15 minutes to 3 minutes in 3-minute increments, while maintaining the complete set of 40 matrix sentences. This reduction impacts only the training data for the decoders, as all testing is performed only on the matrix \gls{eeg} task data.

\subsubsection{Performance comparison with other papers}

To be able to compare the model performance with the performance obtained in other papers, we propose using the \gls{nrmse} as a metric, as it standardizes the variability of the different behavioral \gls{srt} distributions across papers:

\begin{equation}
\mathrm{NRMSE} = \frac{\sqrt{\frac{1}{N} \sum_{i=1}^{N} \left(y_i - \hat{y}_i\right)^2}}{y_{\max} - y_{\min}}
\label{eq:nrmse}
\end{equation}

where \(N\) is the number of observations, \(y_i\) are the behavioral \gls{srt} values, \(\hat{y}_i\) are the predicted \gls{srt}, and \(y_{\max}\) and \(y_{\min}\) are the maximum and minimum behavioral \gls{srt} values, respectively.

To obtain the behavioral and predicted \glspl{srt} from other studies, we extracted data points from published graphs using WebPlotDigitizer (Automeris) \parencite[]{webplotdigitizer}. We were able to retrieve these values from the following papers: \cite{vanthornhoutSpeechIntelligibilityPredicted2018, accouPredictingSpeechIntelligibility2021, borgesSpeechReceptionThreshold2025}.

\subsubsection{SHAP analysis}

\Gls{shap} analysis \parencite[]{lundbergUnifiedApproachInterpreting2017} was used to interpret the contribution of each \gls{erf}-adjusted \gls{nt} value to the prediction of the \gls{srt} outcome. To further explore model interpretability, these \gls{nt} values were grouped by parameter combinations, allowing us to assess the relative importance of each lag, decoder type, stimulus feature, \gls{eeg} task, and frequency band.

\section{Results}

\subsection{SRT prediction}

\subsubsection{Behavioral}
The participants’ average \gls{srt} on the behavioral matrix test  was $-9.07$~\gls{symb:dB}, with a standard deviation of 0.56~\gls{symb:dB}.

\subsection{Model results}

\begin{figure*}[htbp]
  \centering
  \includegraphics[width=0.9\textwidth]{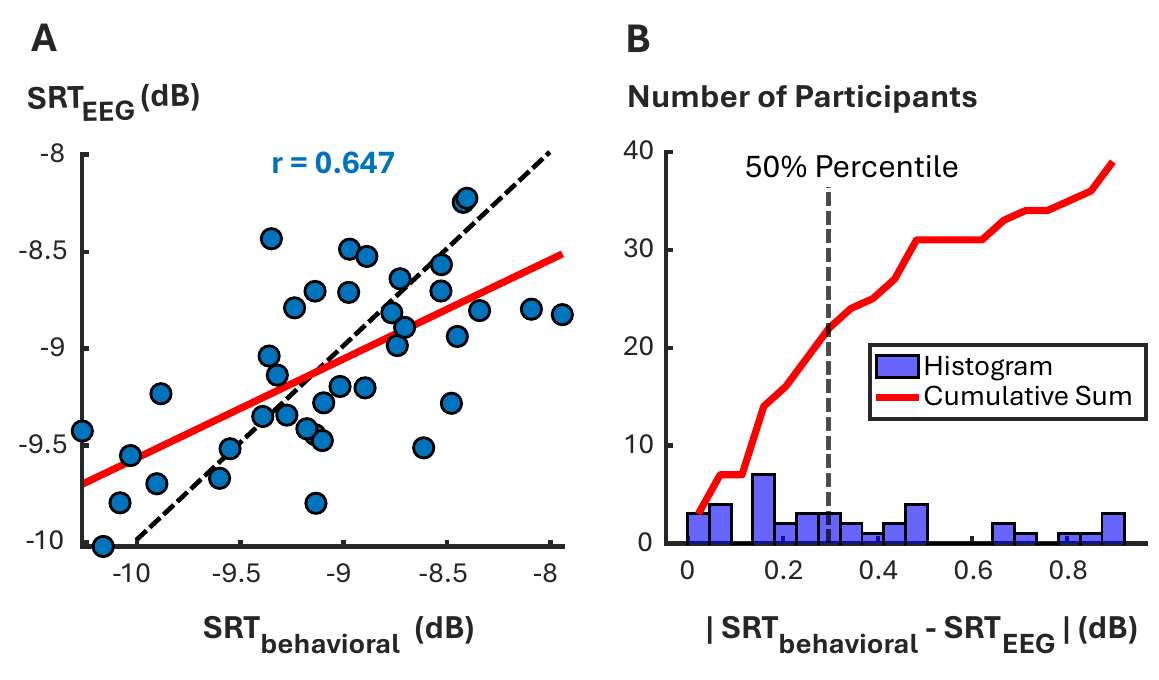}
    \caption[Model performance results]{
    \textbf{Model performance results.} (A) Correlation ($p < 0.001$) between  the behavioral \gls{srt} (\gls{srt}$_{behavioral}$) and the predicted \gls{srt} (\gls{srt}$_{eeg}$). The blue points are subjects, the red line is the regression fit, and the dotted line is the identity line. (B) Histogram of the absolute differences between \gls{srt}$_{behavioral}$ and \gls{srt}$_{eeg}$.}
  \label{fig:fig4}
\end{figure*}

The \gls{svr} model, trained to learn a linear mapping between the \gls{erf}-adjusted \gls{nt} vectors and behavioral \glspl{srt}, achieved a Pearson correlation of 0.647 ($p < 0.001$) between predicted \glspl{srt} and behavioral \glspl{srt} (see \textit{Figure~\ref{fig:fig4}A}). The median absolute difference between predicted and behavioral \glspl{srt} was 0.29~\gls{symb:dB}, with a maximum difference of 0.91~\gls{symb:dB}, as shown in \textit{Figure~\ref{fig:fig4}B}.

\subsection{SHAP results}

\begin{figure}[htbp!]
  \centering
  \includegraphics[width=0.49\textwidth]{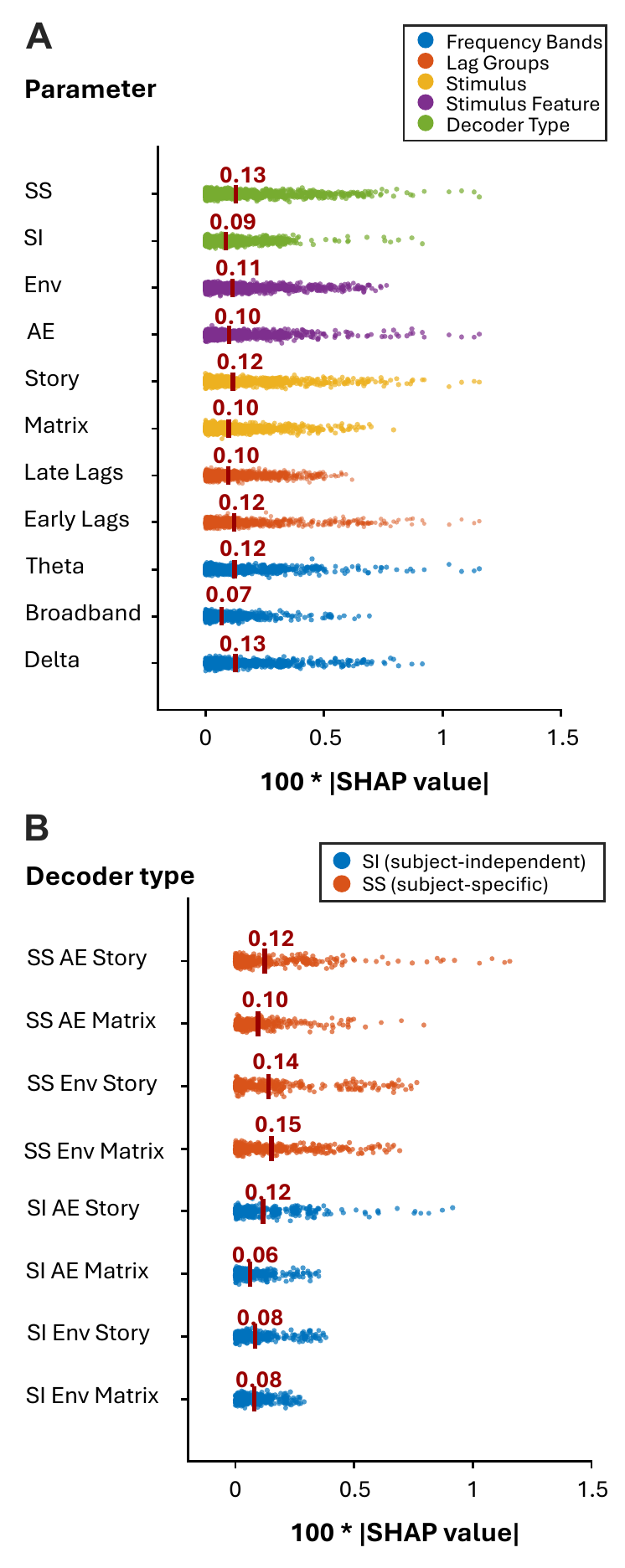}
    \caption[SHAP analysis results]{
    \textbf{SHAP analysis results.} Each point represents the \gls{shap} value of a \gls{erf}-adjusted \gls{nt} value for a specific parameter used in the decoder configuration. \gls{shap} values are grouped as follows: (A) \gls{shap} values are organized by parameter group. If a SHAP value corresponds to multiple parameters (e.g., both \gls{ss} and \gls{env}), it appears in each relevant parameter group. (B) \gls{shap} values are grouped according to the \gls{ss} and \gls{si} decoders type.}
  \label{fig:fig5}
\end{figure}

From the \gls{shap} analysis (\textit{Figure~\ref{fig:fig5}A-B}) it appears that while all parameters contribute, theta/delta bands, early lags, story stimulus, and the \gls{ss} decoder have the strongest influence on model predictions. 

\subsection{Data reduction results}

\begin{figure*}[htbp!]
  \centering
  \includegraphics[width=1.0\textwidth]{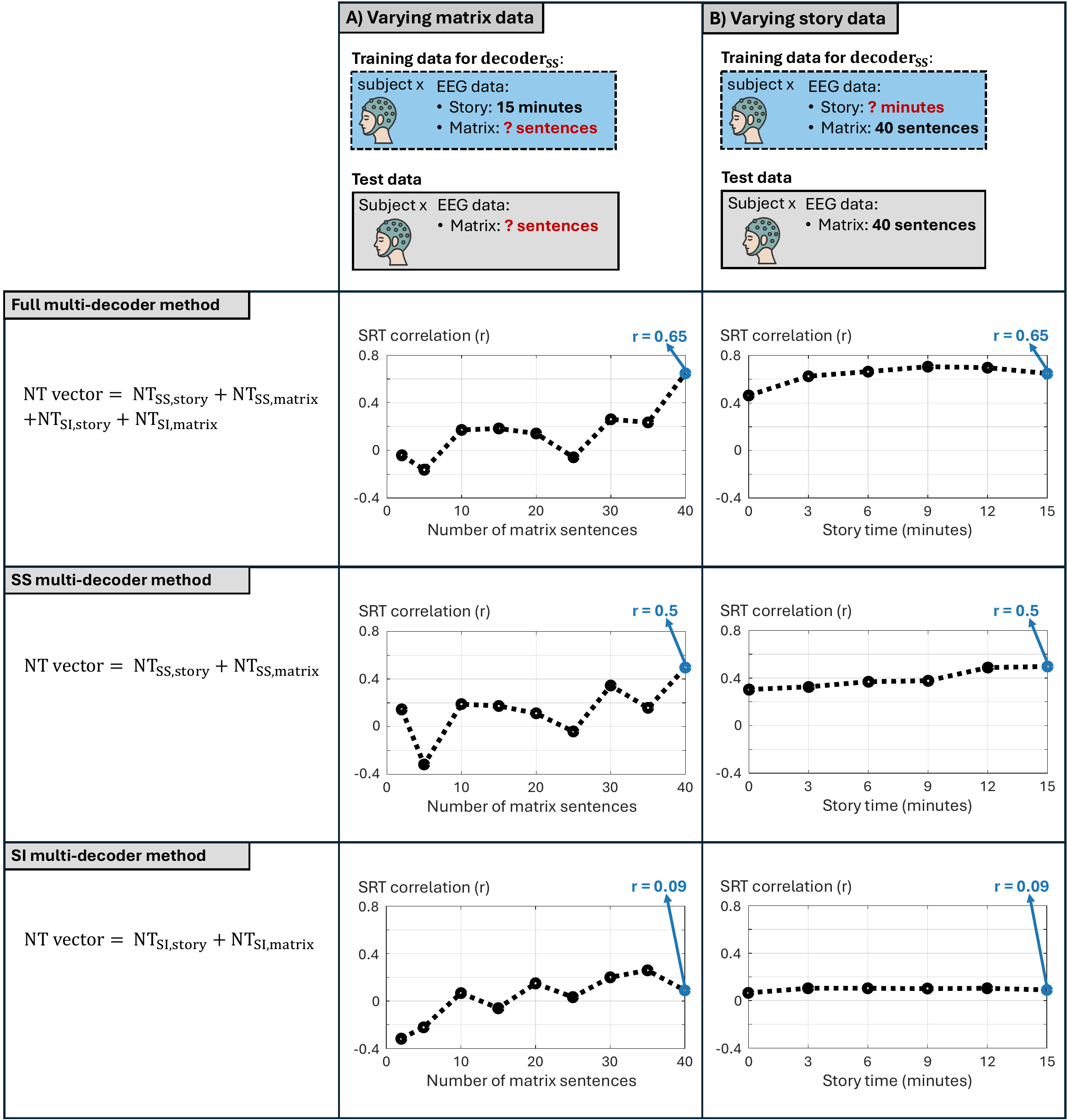}
    \caption[EEG data reduction results]{
    \textbf{EEG data reduction results.} (A) Illustrates the effect of reducing the amount of matrix \gls{eeg} data, measured by the number of matrix sentences available to the model, while keeping the story \gls{eeg} data constant at 15 minutes. The number of sentences cannot fall below 2, as the \gls{ss} decoder requires at least one sentence for training and one sentence for testing. In the full multi-decoder method, the \gls{nt} feature vector is constructed using outputs from both the \gls{si} and \gls{ss} decoders. The \gls{ss} multi-decoder method uses only \gls{ss} decoders, whereas the \gls{si} multi-decoder method employs only \gls{si} decoders. (B) Shows the impact of reducing the amount of story \gls{eeg} data on model performance, while keeping the number of matrix sentences constant at 40.}
  \label{fig:fig6}
\end{figure*}

The data reduction analysis reveals that decreasing the number of matrix sentences in the matrix \gls{eeg} task leads to an immediate decline in model performance, as shown in \textit{Figure~\ref{fig:fig6}A}. In contrast, reducing the duration of the story \gls{eeg} task does not affect performance; the relatively flat line indicates that model accuracy remains stable when the story \gls{eeg} data is reduced from 15 to 3 minutes, see \textit{Figure~\ref{fig:fig6}B}. \textit{Figure~\ref{fig:fig6}} also demonstrates that combining both the \gls{si} and \gls{ss} decoders is necessary to achieve the best model performance, outperforming the use of either decoder individually.

\subsection{Comparison model results with other papers}

\textit{Table~\ref{tab:study_performance}} summarizes metrics from various studies. Although the current study does not demonstrate the highest correlation, it achieves the lowest \gls{nrmse} of 0.19 when the dataset variability is taken into account.

\begin{table*}[htbp]
\centering
\caption{Summary of behavioral and model performance metrics across studies, including mean and standard deviation (SD) of speech reception thresholds (SRT), correlation coefficients, median and standard deviation of the absolute dB differences, and normalized root mean square error (NRMSE). In this table, Sonck et al., 2026 refers to the current study.}
\vspace{0.3cm}
\label{tab:study_performance}
\begin{tabular}{llcccccc}
\toprule
\textbf{N} & \textbf{Study} & \textbf{Mean SRT} & \textbf{SD SRT} & \textbf{Correlation} & \textbf{Median dB Diff.} & \textbf{SD dB Diff.} & \textbf{NRMSE} \\
\midrule
24 & \cite{vanthornhoutSpeechIntelligibilityPredicted2018} & -7.6 & 1.48 & 0.69 & 1.7 & 3.17 & 0.86 \\
19 & \cite{lesenfantsPredictingIndividualSpeech2019a} &  & 0.95 &  &  &  &  \\
18 & \cite{munckePredictionSpeechIntelligibility2022} &  & 1.2 &  &  &  &  \\
20 & \cite{accouDecodingSpeechEnvelope2023} & -8.73 & 0.83 & 0.59 & 3.64 & 1.68 & 1.26 \\
22 & \cite{borgesSpeechReceptionThreshold2025} & -5.34 & 0.6 &  & 0.38 & 1.45 & 0.52 \\
39 & Sonck et al., 2026 & -9.07 & 0.56 & 0.65 & 0.29 & 0.91 & 0.19 \\
\bottomrule
\end{tabular}
\end{table*}

\section{Discussion}

We present a new method, the multi-decoder method, for predicting the \gls{srt} from subjects’ \gls{eeg} recordings obtained while listening to both matrix sentences and a continuous story. Central to this method is the introduction of \gls{erf}-adjusted \gls{nt} vectors, these vectors captures \gls{nt} information across a diverse set of decoder configurations. 

Specifically, the \gls{nt} vectors were constructed by varying key decoder parameters, including the \gls{eeg} task and thus speech stimulus (story vs. matrix), the speech feature (envelope vs. acoustic onsets), the post-integration window (up to 500 \gls{symb:ms}), frequency bands (theta, delta, and broadband) and the type of decoder (subject-independent vs. subject-specific). This resulted in high-dimensional \gls{nt} vectors. After noise-level subtraction and transformation with the error function, the resulting \gls{erf}-adjusted \gls{nt} vectors were used in a \gls{svr} model to predict individual behavioral \glspl{srt}.

Using this model, we successfully predicted the \gls{srt} for all 39 participants, achieving a significant Pearson correlation of 0.647 between the predicted \gls{eeg}-based \gls{srt} and the behavioral \gls{srt}. The median absolute difference between the predicted and behavioral \glspl{srt} across participants was 0.29~\gls{symb:dB} with a standard deviation of 0.26~\gls{symb:dB}, with this difference remaining below 1~\gls{symb:dB} for every individual.

The primary innovation of our approach lies in the integration of information from a wide array of decoders and feature sets. In contrast, previous methods typically relied on a single decoder type and feature set \parencite[]{vanthornhoutSpeechIntelligibilityPredicted2018,lesenfantsPredictingIndividualSpeech2019a, borgesSpeechReceptionThreshold2025,munckePredictionSpeechIntelligibility2022}. Our method eliminates the need for sigmoid fitting, using a \gls{svr} instead. This change overcomes the limitations of some of the earlier methods, which often struggled to fit all subjects. \textcite{vanthornhoutSpeechIntelligibilityPredicted2018} could not fit 21\% of the subjects (5 out of 24). \textcite{borgesSpeechReceptionThreshold2025} could not fit 9\% of the subjects (2 out of 22), \textcite{accouPredictingSpeechIntelligibility2021} could not fit 20\% (4 out of 20) of the subjects and \textcite{lesenfantsPredictingIndividualSpeech2019a} could on average not fit 17\% of the participants. 	However, both \textcite{lesenfantsPredictingIndividualSpeech2019a} and \textcite{munckePredictionSpeechIntelligibility2022} proposed alternative methods that bypass sigmoid fitting, enabling \gls{srt} prediction for all participants. \textcite{lesenfantsPredictingIndividualSpeech2019a} combined time-aligned phonetic features with the spectrogram \parencite[]{dilibertoLowFrequencyCorticalEntrainment2015}, while \textcite{munckePredictionSpeechIntelligibility2022} applied a time-windowed root mean square approach. 

When comparing \gls{eeg}-based models for predicting behavioral \glspl{srt}, it is important to consider the accuracy of the behavioral \glspl{srt} itself. Good quality speech materials  has a behavioral test–retest differences are typically around 0.4~\gls{symb:dB} (median) with a standard deviation of 0.7~\gls{symb:dB} \parencite[]{boonMulticenterEvaluatieValidatie2014,decruySelfAssessedBekesyProcedure2018}.
A challenge in comparing model performance across studies arises from the variability in behavioral \gls{srt} data between datasets, as illustrated in \textit{Table~\ref{tab:study_performance}}. Studies with lower \gls{srt} variability, including our own, may have an inherent advantage in achieving lower median dB differences between behavioral and predicted \gls{srt}. For example, a model that always predicts the average \gls{srt} would yield a lower median dB difference on datasets with smaller standard deviations compared to those with higher variability. To enable fairer comparisons, we recommend adopting metrics that normalize for dataset variability, such as the \gls{nrmse}. While previous studies did not report the \gls{nrmse} metric directly, we applied it retrospectively to their published results for a standardized comparison. These recalculated values are 0.86 \parencite{vanthornhoutSpeechIntelligibilityPredicted2018}, 0.52 \parencite{borgesSpeechReceptionThreshold2025}, and 1.26 \parencite{accouPredictingSpeechIntelligibility2021}. In contrast, our multi-decoder approach achieves a substantially lower \gls{nrmse} of 0.19.

The \gls{shap} analysis of the \gls{svr} model demonstrated that all parameters contribute to overall model performance; however, certain groups—specifically the theta and delta frequency bands, early lags, the story stimulus, and the \gls{ss} decoder—exert a somewhat greater influence on the model’s predictions. The prominence of early lags is not unexpected, as both speech features used in this study, envelope and acoustic onsets, are primarily bottom-up processed and less susceptible to top-down effects such as priming \parencite[]{karunathilakeNeuralTrackingMeasures2023b}. Both delta and theta bands play a crucial role in speech understanding, reflecting the brain’s hierarchical processing of linguistic units such as sentences, phrases, words, syllables, and phonemes \parencite[]{dingCorticalTrackingHierarchical2016}. This aligns with the modulation spectrum of natural speech, which contains prominent energy in the delta (1–4~\gls{symb:Hz}) and theta (4–8~\gls{symb:Hz}) ranges \parencite[]{varnetCrosslinguisticStudySpeech2017}. 

This is further supported by previous findings, where the delta band was most predictive in \cite{vanthornhoutSpeechIntelligibilityPredicted2018}, while the theta band yielded the best results in \cite{lesenfantsPredictingIndividualSpeech2019a}.
An interesting pattern emerges when comparing decoder types: \gls{si} decoders consistently achieve higher \gls{nt} values across \glspl{snr}. This suggests that their advantage may be due to having access to a larger amount of training data, and that it is possible that \gls{ss} decoders would reach similar \gls{nt} values if provided with equivalent amount of data. In contrast, \gls{ss} decoders yield higher \gls{shap} values in the model, highlighting their greater contribution to predicting individual behavioral outcomes. This is further supported by the results that a multi-decoder method using only \gls{ss} decoders performs much better than a multi-decoder method using only \gls{si} decoders.  This suggests that \gls{ss} decoders are better at capturing unique, individual-specific neural features that are particularly informative for predicting each subject’s \gls{srt}.

In the present study, we initially used 29 minutes of \gls{eeg} data per subject. However, since the silence condition in the matrix \gls{eeg} task was excluded, the effective recording time per participant was reduced to 27 minutes. Our time reduction analysis demonstrated that the duration of the story \gls{eeg} task can be shortened from 15 to 3 minutes without compromising model performance, thereby reducing the total required \gls{eeg} recording time to approximately 15 minutes. Conversely, reducing the number of matrix sentences in the matrix \gls{eeg} task is not advisable, as this led to a decline in model accuracy. Consequently, when pretrained \gls{si} decoders are available from this dataset, it is feasible to predict a new participant’s \gls{srt} with only 15 minutes of \gls{eeg} recording. Although this duration exceeds the time needed for a standard behavioral \gls{srt} test, it enhances feasibility for testing populations unable to perform behavioral assessments.

Thus, our proposed method shows potential for clinical application, particularly in populations where active participation is limited or not possible and support the evaluation of various hearing aid strategies. To further improve ecological validity, future research could omit matrix sentences and instead rely entirely on audiobook material presented at different \gls{snr} levels to predict \glspl{srt}, as suggested by \textcite{borgesSpeechReceptionThreshold2025}. Additionally, validating the multi-decoder method in individuals with hearing impairment and older adults would establish its generalizability across a wider range of \glspl{srt}, as the current findings are based solely on participants with normal hearing.

\section{Conclusion}
In this study, we introduced a the multi-decoder method, an \gls{eeg}-based approach for predicting \glspl{srt} using \gls{erf}-adjusted \gls{nt} vectors derived from a diverse set of decoders and with different decoder configurations. This approach enabled accurate \gls{srt} predictions for all participants resulting in a \gls{nrmse} of 0.19.  \gls{shap} analysis indicated that, while all parameters contributed to model performance, certain groups—specifically the theta and delta frequency bands, early lags, and \gls{ss} decoders—had a somewhat greater influence. These results underscore the clinical potential of this method, particularly for populations where behavioral testing is difficult or impractical. Furthermore, we demonstrated that leveraging pretrained \gls{si} decoders allows reducing \gls{eeg} data collection for new participants to just 15 minutes—comprising 3 minutes of story listening and 12 minutes across six \gls{snr} conditions with 40 matrix sentences each—without sacrificing prediction accuracy. Future research should evaluate and validate the method in more diverse clinical populations, including those with hearing loss.

\section{References}
\printbibliography[heading=none]

\appendix
\renewcommand{\thefigure}{\Alph{figure}1} 
\setcounter{figure}{0} 

 \begin{figure}[!t]
  \centering
  \includegraphics[width=0.5\textwidth]{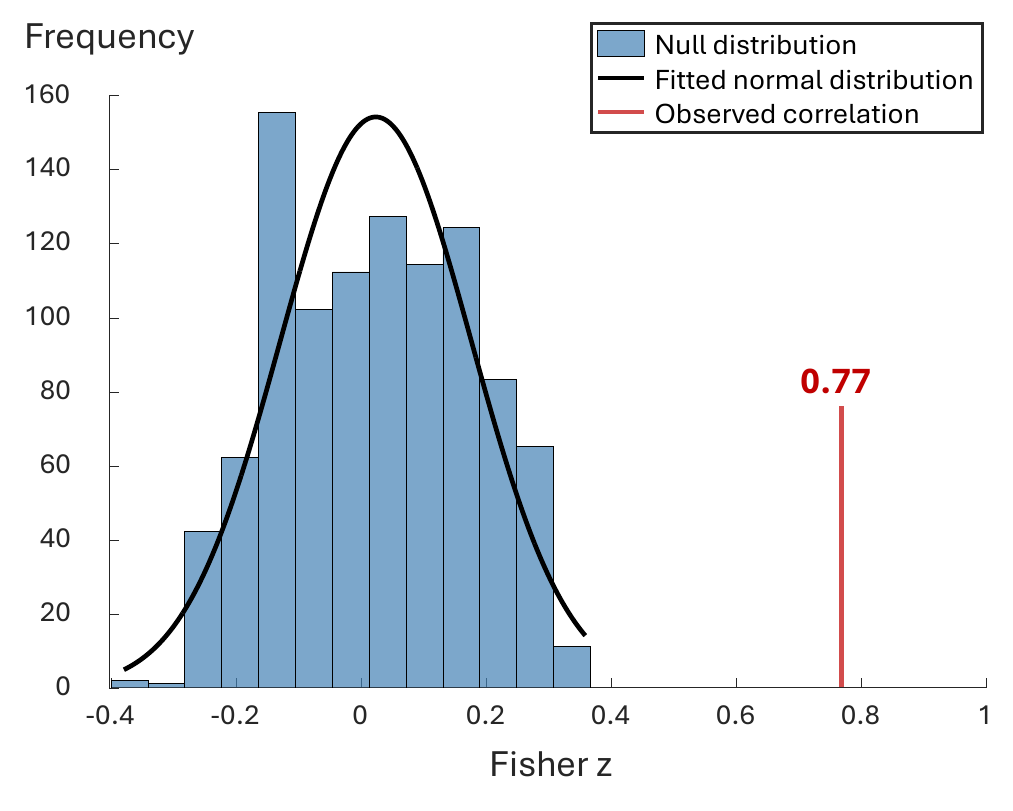}
  \caption{\textbf{Null distribution}. The distribution of Fisher Z-transformed null correlations has a mean of 0.02 and a standard deviation of 0.15. The observed correlation, also Fisher Z-transformed and calculated from the model without disrupting the underlying data structure (indicated by the red line), falls significantly outside this null distribution. This indicates a statistically meaningful relationship beyond chance.}
  \label{fig:figure7}
\end{figure}

\end{document}